\documentclass[reprint,amsmath,amssymb,aps,pra,superscriptaddress,nofootinbib]{revtex4-2}
\usepackage[utf8]{inputenc}
\usepackage[english]{babel}
\usepackage{hyperref}
\usepackage{bm}
\usepackage{hyperref}
\usepackage{physics}
\usepackage{multibib}
\newcites{Method}{References}
\usepackage[dvipsnames]{xcolor}
\usepackage{booktabs}
\usepackage{float}
\usepackage{nicefrac}
\usepackage{graphicx}
\usepackage[capitalise]{cleveref}
\usepackage{mhchem}
\usepackage{booktabs}
\usepackage{tabularx}
\usepackage{ulem}
\usepackage{enumitem}
\usepackage{titlesec}
\usepackage{caption}
\usepackage{circledsteps}

\usepackage{sidecap}

\usepackage[page]{appendix}
\usepackage[
    protrusion=true,
    activate={true,nocompatibiality},
    final,
    tracking=true,
    kerning=true,
    spacing=true,
    factor=1000]{microtype}

\newcolumntype{Y}{>{\centering\arraybackslash}X}

\hypersetup{
 colorlinks=true, 
 linkcolor=blue, 
 citecolor=blue,        
 filecolor=blue,      
 urlcolor=blue}

\newcommand{\beginsupplement}{%
    \renewcommand{\appendixpagename}{\large\centering Supplementary Information: \mytitle}
    \setcounter{table}{0}
    \setcounter{figure}{0}
    \setcounter{equation}{0}
    \renewcommand{\thetable}{S\arabic{table}}%
    \renewcommand{\thefigure}{Supplementary Fig. \arabic{figure}}%
    \renewcommand{\thesection}{S-\arabic{section}}
    \renewcommand{\theequation}{S.\arabic{equation}}
    \onecolumngrid
    \widetext
    \appendix   
    \appendixpage
    
}
\Crefformat{appendix}{Supplementary Information #2#1#3}
\titleformat*{\section}{\bfseries}
\titleformat*{\subsection}{\bfseries}

\definecolor{gregRed}{RGB}{209, 0, 11}

\begin{document}

                                                         
\author{Gr\'egory Moille}
\email{gregory.moille@nist.gov}
\affiliation{Joint Quantum Institute, NIST/University of Maryland, College Park, USA}
\affiliation{Microsystems and Nanotechnology Division, National Institute of Standards and Technology, Gaithersburg, USA}
\author{Usman A. Javid}
\affiliation{Joint Quantum Institute, NIST/University of Maryland, College Park, USA}
\affiliation{Microsystems and Nanotechnology Division, National Institute of Standards and Technology, Gaithersburg, USA}
\author{Michal Chojnacky}
\affiliation{Joint Quantum Institute, NIST/University of Maryland, College Park, USA}
\affiliation{Sensor Science Division, National Institute of Standards and Technology, Gaithersburg, USA}
\author{Pradyoth Shandilya}
\affiliation{University of Maryland at Baltimore County, Baltimore, MD, USA}
\author{Curtis Menyuk}
\affiliation{University of Maryland at Baltimore County, Baltimore, MD, USA}
\author{Kartik Srinivasan}
\email{kartik.srinivasan@nist.gov}
\affiliation{Joint Quantum Institute, NIST/University of Maryland, College Park, USA}
\affiliation{Microsystems and Nanotechnology Division, National Institute of Standards and Technology, Gaithersburg, USA}
\date{\today}

\newcommand{\mytitle}{AC-Josephson Effect and Sub-Comb Mode-Locking in a Kerr-Induced Synchronized Cavity Soliton}
\title{\mytitle}

                                                             
\begin{abstract} 
    Kerr-induced synchronization (KIS)~\cite{MoilleNature2023} involves the capture of a dissipative Kerr soliton (DKS) microcomb~\cite{KippenbergScience2018} tooth by a reference laser injected into the DKS resonator. This phase-locking behavior is described by an Adler equation whose analogous form describes numerous other physical systems~\cite{CoulletAmericanJournalofPhysics2005}, such as Josephson junctions~\cite{JainPhysicsReports1984}. We present an AC version of KIS whose behavior is similar to microwave-driven Josephson junctions, where periodic synchronization occurs as so-called Shapiro steps. We demonstrate consistent results in the AC-KIS dynamics predicted by the Adler model, Lugiato-Lefever equation, and experimental data from a chip-integrated microresonator system. The (integer) Shapiro steps in KIS can simply be explained as the sideband created through the reference laser phase modulation triggering the synchronization. Notably, our optical system allows for easy tuning of the Adler damping parameter, enabling the further observation of fractional-Shapiro steps, where the synchronization happens at a fraction of the driving microwave frequency. Here, we show that the comb tooth is indirectly captured thanks to a four-wave mixing Bragg-scattering process, leading to sub-comb mode-locking, and we demonstrate this experimentally through noise considerations. Our work opens the door to the study of synchronization phenomena in the context of microresonator frequency combs, synthesis of condensed-matter state analogues with DKSs, and the use of the fractional Shapiro steps for flexible and tunable access to the KIS regime.
\end{abstract}
\maketitle

Synchronization phenomena are ubiquitous in nature and have been observed in a wide range of systems~\cite{Strogatz2004sync}. Some examples in the biological sciences include neurons~\cite{FellNatRevNeurosci2011} and the blinking pattern of fireflies~\cite{BuckScience1968}, while in the physical sciences and engineering triode oscillators~\cite{Appleton1922}, damped and driven pendulums~\cite{CoulletAmericanJournalofPhysics2005}, and Josephson junctions (JJs) made of two superconductors separated by an insulator~\cite{JainPhysicsReports1984, VlasovPhys.Rev.E2013} have all been studied. In optics, phase-locking is a well-known phenomenon and has been observed, for instance, with optical injection locking of Fabry-Perot lasers~\cite{Lang_injection_1982}. In the context of chip-integrated nonlinear photonics, synchronization phenomena involving dissipative Kerr solitons (DKSs) in optical microresonators have also been studied, including studies of counterpropagating DKSs within one resonator~\cite{YangNaturePhoton2017} and DKSs within two independent resonators~\cite{Jang_synchronization_2018}. More recently, Kerr induced synchronization (KIS) has been demonstrated~\cite{MoilleNature2023}, where an external reference pump laser captures a DKS comb tooth through the phase-locking of the DKS to the reference intracavity field, thanks to the intrinsic nonlinearity of the cavity where the DKS lives.

Although many of the above systems seem at first unrelated, their synchronization mechanism can be represented with the same two coupled-oscillator model, where phase-locking happens under the right condition. Mathematically, synchronization can be described in the form of the Adler equation~\cite{AdlerProc.IRE1946}. For KIS, the phase locking mechanism between the comb tooth and the injected reference pump also follows an Adler equation of the second order, identical to the JJ in the RCSJ (resistively and capacitively shunted junction) model~\cite{McCumberJ.Appl.Phys.1968}. This remarkable similarity suggests the opportunity to leverage discoveries made in one system and apply them to the other. In particular, JJs are of special interest in their (reversed) AC regime~\cite{JosephsonPhysicsLetters1962a}, where the junction is bathed in a microwave field. This system presents periodic synchronization regimes within the parameter space of voltage against current in the junction, a phenomenon known as Shapiro steps~\cite{ShapiroPhys.Rev.Lett.1963}. This period is defined by the microwave frequency factored by $\hbar/2e$ ($\hbar$ is Planck's constant divided by 2$\pi$ and $e$ is the electron's charge), enabling the Volt to be defined based on fundamental constants within the International System of Units (SI)~\cite{FieldMetrologia1973}. Moreover, fractional Shapiro steps have been observed in different junction schemes~\cite{RokhinsonNaturePhys2012, HuangAppliedPhysicsLetters2023, VignaudNature2023}, and are created through the presence of higher harmonics or the second-order nature of the relevant Adler equation model~\cite{ValizadehJNMP2008}. Given that KIS is also governed by a second-order Adler model, we predict that similar effects should be present when studying its AC regime, which we access by introducing a phase modulation in the reference pump laser. Our experimental platform, a chip-integrated Kerr nonlinear microresonator that generates an octave-spanning DKS comb, exhibits both integer and fractional Shapiro steps. While the integer Shapiro steps are straightforward to understand through sideband synchronization, the fractional Shapiro steps have a more subtle explanation. We show that the effective damping of the two-oscillator model dictates their presence, which can be explained in the context of four-wave mixing Bragg-scattering (FWM-BS), resulting in the mode-locking of the sub-comb formed by the nonlinear interaction between the comb tooth and the reference pump. The fractional Shapiro steps exhibit an indirect capture of the comb tooth that is particularly of interest for tunability of the KIS regime. From noise measurements, we confirm the mode-locking through FWM-BS by demonstrating that the beat note between the reference pump and the DKS comb tooth does not carry any optical noise from the lasers and only exhibits noise from the microwave generator used to apply phase modulation. Our experimental observations and physical explanations are corroborated by Lugiato Lefever equation (LLE) and Adler equation simulations, further highlighting how the detailed nonlinear physics describing our photonic system can be mapped onto a universal synchronization model. 

\begin{figure*}[!t]
    \begin{center}
        \includegraphics{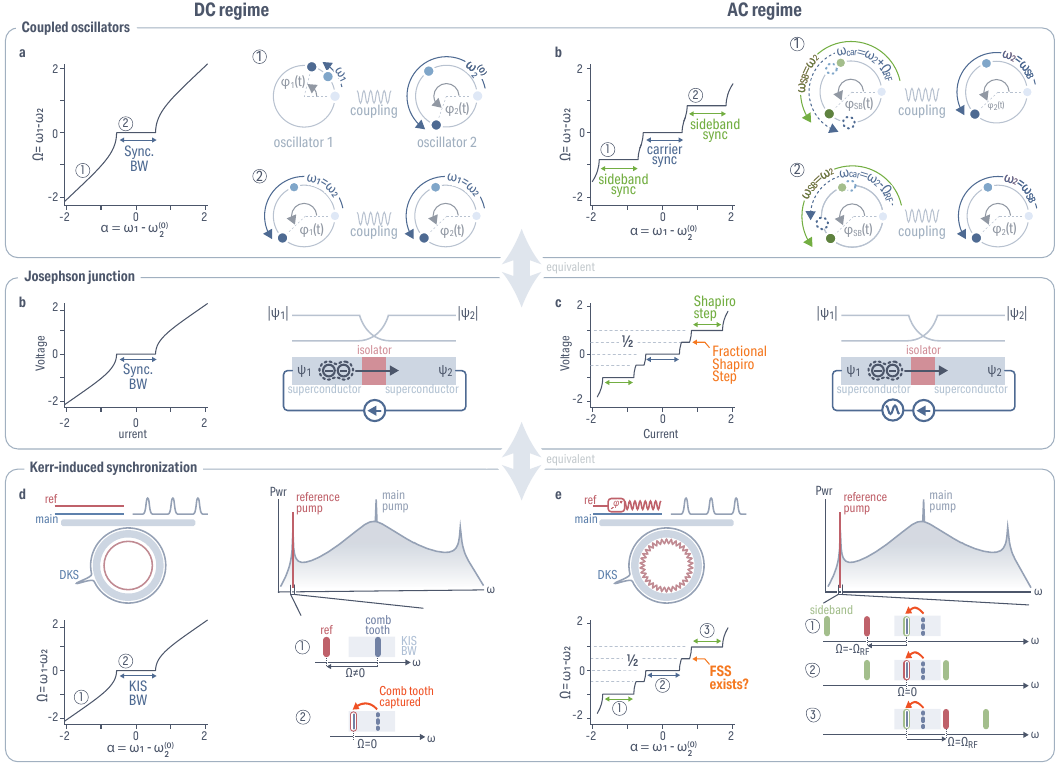}
    \end{center}
    \caption{\label{fig:1}
        \textbf{Concept of the DC and AC synchronization regimes and equivalence between systems -- } %
        \textbf{a} Two oscillators, one at a controllable rotation speed $\omega_1$ (left) and the other at a fixed original rotation speed $\omega_2^{(0)}$, are coupled together. If $\omega_1$ is too far from $\omega_2^{(0)}$, they will rotate independently (\Circled{1}). However, if $\omega_1$ is close enough to $\omega_2^{(0)}$, synchronization happens (\Circled{2}), leading to phase-locking of both oscillators with $\Omega = \omega_1 - \omega_2 = 0$ (hence $\omega_2$ can be different from $\omega_2^{(0)}$). %
        \textbf{b} In the AC regime, $\omega_1$ is modulated at $\Omega_\mathrm{RF}$, here being unity for simplicity.
        If the rotation speed of the controllable oscillator is close enough to the fixed one, synchronization happens, either with the trailing speed, resulting in lower side-band synchronization (\Circled{1}), or with the leading one for higher side-band synchronization (\Circled{2}). %
        \textbf{c} Josephson junctions (JJs) in their macroscopic description are equivalent to the coupled oscillator since they follow the same Adler equation. Cooper-pairs in adjacent superconductors couple thanks to overlapping wavefunction amplitudes, synchronizing their phase, and thus creating observable features in the junction's current/voltage space.
        \textbf{d} In the AC-regime, JJs exhibit integer Shapiro steps (ISSs) in current/voltage, useful for defining the Volt standard. Importantly, under certain conditions, fractional Shapiro steps (FSSs) can also occur.
        \textbf{e} In the context of Kerr-induced synchronization (KIS), a dissipative Kerr soliton (DKS) is created in a microring resonator. Injecting a reference pump can lead to the capture of a comb tooth by the reference. The dynamics are similar to \textbf{a} and \textbf{c}, where the follower oscillator is the DKS.  %
        \textbf{f} Like b-d, an AC-KIS regime should cause Shapiro steps via phase modulation of the reference. The ISS is intuitively understood by treating the reference's modulation sidebands as independent lasers that can trigger KIS. While the FSS lacks a simple KIS explanation, the KIS system's tunability should make them observable.
}
\end{figure*}

\begin{figure*}[t]
    \begin{center}
        \includegraphics{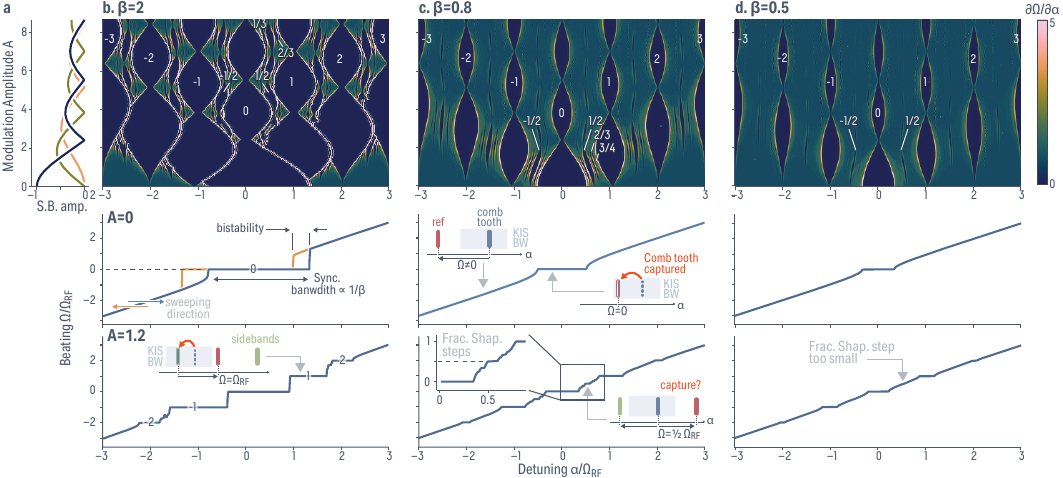}
        \caption{\label{fig:2}
        \textbf{Impact of damping on the synchronization phase space in the AC-KIS regime -- } %
        \textbf{a} Bessel functions which depict the strength of the carrier (black), first sidebands (green) and second sidebands (yellow) in the context of an optical phase modulator. %
        \textbf{b-d} Top: Solutions to the Adler equation which map out the synchronization phase space of the AC-KIS regime for different damping parameter $\beta =2, 0.8$ and 0.5, respectively (under-damped, balanced, over-damped), while the normalized RF frequency is $\Omega_\mathrm{RF} = 1.5$. The color-maps correspond to the ``effective resistivity'' $\partial\Omega/\partial\alpha$, similar to the JJ synchronization phase space. The zeros of the Bessel functions match the bottlenecks of the corresponding synchronization windows. Bottom: Synchronization for the DC case (\textit{i.e.} A = 0) and for a selected modulation amplitude strength. Integer Shapiro steps are highlighted by the $\pm 1$ and $\pm 2$ labels for their corresponding sideband KIS. In the case of $\beta=0.8$, the fractional Shapiro steps are clearly visible, and do not correspond to any direct comb tooth capture. %
        }
    \end{center}
\end{figure*}


\vspace{1em}
\textbf{Theoretical model --} We recall the physical intuition behind the synchronization, first through its already introduced DC regime, and then we will introduce its AC regime. In the DC regime the second-order Adler equation describing synchronization can be normalized such that: 

\begin{equation}
    \beta \frac{\partial^2 \Phi}{\partial \tau^2} + \frac{\partial \Phi}{\partial \tau} = \alpha + \sin(\Phi)
    \label{eq:DC_Adler}
\end{equation}

\noindent with $\tau = \Omega_0 t$ being the normalized time with respect to the oscillators' synchronization bandwidth $2\Omega_0$,  $\beta = \nicefrac{\Omega_0}{\Gamma}$ the damping factor with $1/\Gamma$ the characteristic energy loading time of the system, $\Phi = \phi_1 -\phi_2$ the difference of phase between the two oscillators, $\partial \Phi /\partial \tau = \omega_1 - \omega_2 = \Omega $ the difference of rotation frequencies between the two oscillators, and $\alpha = \nicefrac{\left(\omega_1 - \omega_2^{(0)} \right)}{\Omega_0}$ the normalized detuning relative to the original rotation speed of the follower oscillator $\omega_2^{(0)}$, acting as an effective loss term in the differential equation. For a comparison of the physical significance of these parameters for different systems, including the microcomb one under study, we refer the reader to~\cref{sec:sup_parameter}. In many cases -- for instance with fireflies blinking, two coupled pendulums, or overdamped JJs -- the second-order derivative can be neglected since $\beta \ll 1$. However, in the context of KIS, this parameter plays an important role which we shall study further. When the two oscillators present frequencies that are too far apart, they will oscillate independently [\cref{fig:1}a]. However, if the frequency difference is within the locking range $\Delta \alpha = \pm 1$, synchronization happens, defined by $\nicefrac{\partial \Phi }{\partial \tau} = 0 $. Here, the two oscillators will be phase-locked and hence rotate at the same frequency. A similar phenomenon happens with the cooper pairs in the JJ [\cref{fig:1}c] and in KIS [\cref{fig:1}e].

The AC regime assumes the introduction of a periodic modulation of the phase in one of the oscillators~[\cref{fig:1}b] such that: 
\begin{equation}
    \beta\frac{\partial^2 \Phi}{\partial \tau^2} + \frac{\partial \Phi}{\partial \tau} = \alpha + \sin(\Phi) + A\sin(\Omega_\mathrm{RF}\tau)
    \label{eq:AC_Adler}
\end{equation}
where $A$ is the modulation amplitude and $\Omega_\mathrm{RF} = \Omega_0^{-1}\widetilde{\Omega_\mathrm{RF}}$ is the normalized phase modulation modulation. In this regime, the synchronization still happens when $\Delta \alpha = \pm 1$; however, new synchronization windows also appear. In a temporal representation, the reference oscillator may be trailing (leading) the follower one. If the reference is modulated, its oscillation introduces another leading (trailing) oscillation [\cref{fig:1}b]. If the rotation frequency of the reference is within the locking range minus (plus) the modulation frequency, the modulation will be sufficient to enter the synchronization regime. This phenomenon is of particular interest in the context of JJs, where the synchronization can be represented in the voltage versus current space [\cref{fig:1}d], yielding a macroscopic description of the aggregate effects associated with quantum systems (Cooper pairs of the two insulator-gapped  superconductors) and phase locking in the same fashion as the previously introduced coupled oscillators. In the AC regime, new synchronization windows are also created, the so-called integer Shapiro steps (ISS), at fixed voltage plateaus [\cref{fig:1}d]. This direct relationship between voltage and current through physical constants and applied microwave frequency is at the heart of primary voltage standards~\cite{FieldMetrologia1973,TaylorMetrologia1989, StockMetrologia2019}. Interestingly, under the right conditions synchronization can also occur at so-called fractional Shapiro steps (FSS), providing a unique tool to study very rich physics~\cite{PientkaPhys.Rev.X2017, RokhinsonNaturePhys2012, VignaudNature2023, VlasovPhys.Rev.E2013}.

Optically, we harness the recently introduced KIS to study the physics of \cref{eq:DC_Adler,eq:AC_Adler}. A DKS is first generated in a whispering gallery mode resonator, here depicted as a microring, thanks to the nonlinearity/dispersion and loss/pumping double-balance~\cite{KippenbergScience2018}. The DKS being periodically extracted by an access waveguide results in a pulse train obtained at the output of the system, which through the Fourier relationship results in a frequency comb that can span up to or beyond an octave~\cite{LiOptica2017a, PfeifferOpticaOPTICA2017, MoilleNat.Commun.2021a}. Then, a reference pump laser is injected in the same cavity [\cref{fig:1}e]. Thanks to the same nonlinearity generating the DKS, the reference pump can capture the nearest comb tooth if close enough in frequency, resulting in KIS of the DKS microcomb~\cite{MoilleNature2023}.  %
One could derive a mathematical model for the KIS dynamics focusing on the modification of the soliton envelope momentum by the reference~\cite{TaheriEur.Phys.J.D2017,WildiAPLPhotonics2023}, which accurately captures the entrainment and optical frequency division of the reference pump onto the DKS repetition rate. This approach can be considered as an extension of the trapping of a DKS through inhomogeneity~\cite{ErkintaloJ.R.Soc.N.Z.2022}, often created through either phase modulation~\cite{JangNatCommun2015} or amplitude modulation~\cite{HendryPhys.Rev.A2018} of the intracavity field at the repetition rate of the DKS, which has been shown to find application in repetition rate disciplining~\cite{WengPhys.Rev.Lett.2019} and synthetic frequency lattices~\cite{EnglebertNat.Phys.2023}. 
An interesting alternative approach for understanding soliton synchronization is to work in the phase velocity soliton space, which allows the effect to be written as an Adler equation~\cite{YangNaturePhoton2017,MoilleNature2023}. In comparison to the DKS momentum approach, the main advantage of this Adler equation approach is that it lets us work with the two time scales of interest that are very far apart: the fast oscillation of the reference pump trapping potential and the phase modulation in the AC-regime, with the former close to two orders of magnitude times faster than the round trip time while the latter is four to five orders of magnitude slower than the round trip time, as the Adler equation is normalized to the natural KIS frequency $\Omega_0= 2\mu_s D_1 E_\mathrm{kis}$. The normalized parameters of~\cref{eq:AC_Adler} become $\tau = t\Omega_0$, $\beta =\nicefrac{\Omega_0}{\kappa}$, $\alpha=(\delta\omega_\mathrm{ref} + D_\mathrm{int}(\mu_s) + \delta\omega_\mathrm{NL})\Omega_0^{-1}$ and $\Phi = \varphi_\mathrm{ref} - \varphi_\mathrm{dks}$. Here, $D_1$ is the DKS repetition rate, $\mu_s$ is the mode normalized to the main pump and is where KIS occurs, $D_\mathrm{int}(\mu_s)$ is the integrated dispersion at the KIS mode, $\delta\omega_\mathrm{NL}$ is the nonlinear frequency shift of the KIS mode from cross-phase modulation, $\delta\omega_\mathrm{ref}$ is the reference detuning, $E_\mathrm{kis} = \sqrt{\frac{\kappa_\mathrm{ext}}{\kappa^2}P_\mathrm{ref}E_\mathrm{\mu s}}/E_\mathrm{dks}$ is the available KIS power and is the geometrical mean of the intracavity reference and comb tooth energy normalized to the DKS energy $E_\mathrm{dks}=\int |a(\theta)|^2 \mathrm{d}\theta/2\pi$, with $a(\theta)$ being the soliton intracavity envelope field, and $\kappa$ and $\kappa_\mathrm{ext}$ are the total and external loss rate. %
Physically, if the continuous wave (CW) reference frequency is too far from the comb tooth, the nonlinearity is not sufficient to pull the comb tooth into synchronization, while the reference creates its own wavepacket traveling at the same group velocity as the soliton thanks to cross-phase modulation coupling~\cite{WangOptica2017a}. Therefore, there is only a discrepancy  of the phase velocity between the two intra-cavity ``colors'' $\nicefrac{\partial \varphi_\mathrm{ref}}{\partial \tau} - \nicefrac{\partial\varphi_\mathrm{dks} }{\partial \tau }\neq 0$ leading to a non-null carrier envelope offset (CEO)   $\Omega = \nicefrac{\partial \varphi_\mathrm{ref}}{\partial \tau} - \nicefrac{\partial\varphi_\mathrm{dks}}{\partial \tau} = \omega_\mathrm{0,ref} \neq 0$. In practice, this CEO offset results in a non-zero beat note between the reference pump and its nearest DKS comb tooth. However, if the reference laser is close enough to one comb tooth, synchronization happens with $\Omega = 0$, corresponding to the carrier envelope offset (CEO) of the DKS aligning with the one of the intracavity reference colors. Experimentally, an absence of the reference/comb tooth beat note can be measured, highlighting the capture of the comb tooth by the reference pump~\cite{MoilleNature2023}. One particular aspect of KIS is the tunable Adler damping parameter $\beta$ in \cref{eq:DC_Adler,eq:AC_Adler}, which is directly related to the reference pump power and allows for simple tuning of the system regime via an externally controllable parameter. 

Optical phase modulators are essential tools in optical experiments and can be obtained as off-the-shelf components that harness the second-order nonlinearity $\chi^{(2)}$ of an appropriate material, enabling electro-optic phase modulation of the reference laser to transform~\cref{eq:DC_Adler} into its AC counterpart in~\cref{eq:AC_Adler} with a simple experimental setup modification [\cref{fig:1}d]. However, contrary to JJs where the microwave bath does not impact the Cooper pair strength, using an optical phase modulator does change the carrier amplitude as energy is transferred to the sideband following a Bessel function dependence, changing~\cref{eq:AC_Adler} to: 
\begin{equation}
    \beta\frac{\partial^2 \Phi}{\partial \tau^2} + \frac{\partial \Phi}{\partial \tau} = \alpha + \eta_\mathrm{car}\sin(\Phi) + \eta_\mathrm{RF} A \sin(\Omega_\mathrm{RF}\tau)
    \label{eq:AC_Adler_Bessel}
\end{equation}

\noindent with $\eta_\mathrm{car} = \cos\left[ \sin\left( \Omega_\mathrm{RF}\tau \right) \right]$ and $\eta_\mathrm{RF}= \cos\left( \Phi \right)$, the carrier and sideband efficiency, respectively, arising from the expansion of the phase modulation term $\sin\left[ \Phi + \sin\left( \Omega_\mathrm{RF} \tau \right) \right]$. It is quite simple to understand the Shapiro steps in the context of the AC-KIS. Phase modulation of a laser is well-known to introduce sidebands in its optical spectrum. Here, the reference laser will be composed of the carrier and two sidebands separated from the carrier by $\pm\Omega_\mathrm{RF}$. It is simple to understand that KIS could occur either for the carrier or for the sidebands, essentially providing a Fourier space picture~[\cref{fig:1}(d)] of the previous temporal explanation provided with the oscillators~[\cref{fig:1}(b)]. To this extent, Shapiro steps in KIS are a straightforward result of the sideband triggering the synchronization, and can be a useful tool for experiments (\textit{i.e.,} a fixed frequency reference laser, which is often the case due to stabilization to an atomic reference, can now be tuned with a phase modulator). However, as we see below, the straightforward ISSs do not encompass the entirety of the physical phenomena observable in AC-KIS, and one could leverage the equivalence between the previously introduced systems to expect FSSs to also be present in the AC-KIS.


\textbf{Theoretical fractional Shapiro steps -- }
As mentioned previously, one particular aspect of KIS is the easily tunable damping parameter $\beta\propto P_\mathrm{ref}^{-1/2}$ while also having direct control over the phase modulation frequency $\Omega_\mathrm{RF}$. Given the modulation sideband strength for an optical phase modulator follows Bessel functions, it is straightforward to assume that the Shapiro step width also follows [\cref{fig:2}a].  The zeroth order synchronization bandwidth from the carrier KIS also changes with the microwave modulation strength, since the parameter $\eta_\mathrm{car}$ accounts for the reduction of the carrier power which is transferred to the sidebands. %
It is particularly interesting to plot the synchronization phase space detuning/beat note against the ``effective resistivity'' $\nicefrac{\partial \Omega}{\partial\alpha}$, similar to well-known JJ synchronization phase space current/voltage. In the under-damped case $\beta=2>\Omega_\mathrm{RF}=1.5$, the KIS bandwidth is larger than the AC modulation frequency, and hence the Shapiro steps overlap [\cref{fig:2}(b)], while as expected their widths follow the Bessel functions.  At large enough modulation strength $A$, other sidebands are created and lead to higher-order Shapiro steps, which are easily understandable by assuming each sideband as a KIS-reference pump. The large contribution of the second-order derivative creates a large bistability, particularly noticeable in the DC regime ($A=0$), which for the AC system could result in an interesting regime where the carrier and sideband KIS become bistable and could potentially find practical use, for instance in optical memories. Here, the direction of the reference laser detuning, controlled through current to the laser, for example, would provide fine control to enable hopping between each Shapiro step. In the over-damped case $\beta = 0.5$~[\cref{fig:2}(d)], the second-order derivative can be mostly neglected, and no bistability is present. Similarly, the Shapiro step widths follow the Bessel function, while $\Omega_\mathrm{RF}\gg \beta$ (\textit{i.e.} $\Omega_\mathrm{RF}$ much larger in comparison to the synchronization bandwidth) makes them clearly separated. This regime is experimentally interesting as this corresponds to low reference pump power, and demonstrates the viability of KIS for repetition rate control and stabilization with $\approx1$~$\mu$W reference pump power levels. Similar behavior as the underdamped regime could also be obtained if one reduces $\Omega_\mathrm{RF}$ to be lower than $\Omega_0$.
More interesting is the in-between regime, neither over- nor under-damped, where $\beta$ is close to unity [\cref{fig:2}(c)]. In this regime, new synchronization arises in the phase space appearing in-between the Shapiro steps, at non-integer rational numbers. These so-called fractional Shapiro steps have been observed in JJs~[\cref{fig:1}(d)], and are a direct consequence of the second-order derivative in the Adler equation~\cite{ValizadehJNMP2008}. In the context of KIS, it is interesting to notice that it leads to an indirect capture of the comb tooth. Essentially, the carrier and sideband reference surround the comb tooth, and as the beat note between the tooth and reference is fixed, repetition rate entrainment of the microcomb still occurs like in the DC-KIS regime. Different FSSs occurs, with, as expected, their width following the well-known Farey sequence~\cite{OdaviccPhys.Rev.E2015, SokoloviccPhys.Rev.E2017}. 
Interestingly, we note that in the underdamped junction, at the zeros of the first-order Bessel function, the reference carrier will be suppressed while only the sidebands are present. The zeroth order synchronization windows from the carrier present a bottleneck shape which instead of completely closing, as is the case for the overdamped junction, is still open as it becomes the $\nicefrac{1}{2}$ FSS of the two first sidebands.

\begin{figure*}
    \begin{center}
        \includegraphics{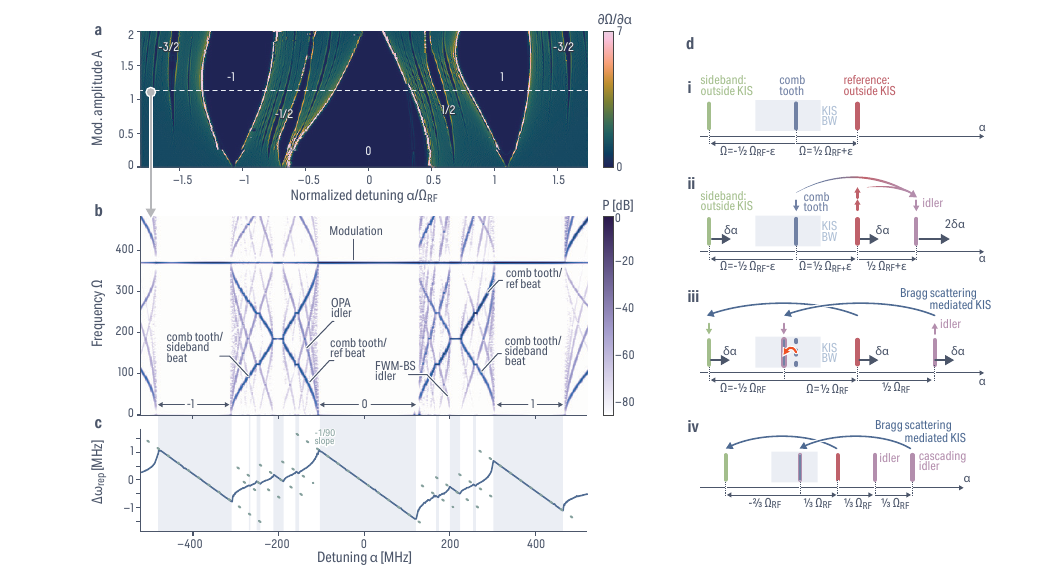}
        \caption{\label{fig:3} 
        \textbf{LLE simulations of the synchronization phase space.} \textbf{a} Phase space of the synchronization between the reference pump and the comb tooth as a function of the modulation strength $A$ and the normalized detuning $\alpha/\Omega_\mathrm{RF}$, with $\widetilde{\Omega_\mathrm{RF}} =372$~MHz. Similar to~\cref{fig:2}b-d, the color-map corresponds to the ``effective resistivity'' $\partial\Omega/\partial\alpha$. Integer (ISSs) and fractional Shapiro steps (FSSs) are highlighted by their respective synchronization order. 
        \textbf{b} Obtained spectrogram (beat note $\Omega$ vs. detuning $\alpha$) of the interference at $\mu_s$ from the four-wave mixing between the carrier, sideband and reference pump. The ISSs are evident as vertical white gaps and correspond to the beat note between the corresponding reference with the comb tooth vanishing. The FSSs are also visible as fixed beat notes (horizontal plateaus) with detuning. Bands exhibiting higher slopes against detuning correspond to different waves arising from the FWM process. 
        \textbf{c} Repetition rate variation $\Delta\omega_\mathrm{rep}$ against $\alpha$, showcasing the $\omega_\mathrm{rep}$ entrainment in every synchronization regime (i.e., for ISSs or FSSs). The dashed lines correspond to the expected optical frequency division (OFD) factor $1/\mu_s = -1/90$, as expected and in good agreement with the simulation. In addition, the FSSs follow the same OFD despite the indirect capture of a comb tooth.
        \textbf{d} Physical explanation of the FSSs and resulting mode locking. (i) In the $\nicefrac{1}{2}$ FSS case, the carrier and sideband surround the comb tooth. (ii) FWM between the carrier (acting as a pump) and sideband (acting as a signal) creates an idler following OPA energy conservation. From energy conservation, any detuning of the carrier results in twice the detuning of the idler, consistent with the spectrogram in \textbf{b}. (iii) The carrier and the sideband act as FWM-BS pumps, enabling frequency translation of any signal by $-\Omega_\mathrm{RF}$.  The OPA idler being at $\Omega_\mathrm{RF}$, it is translated on the comb tooth, triggering KIS and mode locking the system. Any detuning of the carrier is also absorbed by the comb tooth, since it is synchronized through FWM-BS with the detuning of the carrier and OPA idler. (iv) Similar regimes can occur with cascaded OPA idlers, resulting in higher-order FSSs . 
        }
    \end{center}
\end{figure*}
\vspace{1em}

\textbf{Adler equation validation against the Lugiato Lefever Equation -- }
While the Adler model provides an intuitive approach to gaining physical insight about KIS, it has thus far not been demonstrated, either theoretically or experimentally, to accurately capture the KIS dynamics. The AC-KIS regime is a convenient example, in particular with the predicted emergence of FSSs, to verify the validity of the model. First, we aim to reproduce the FSSs using the multi-driven Lugiato Lefever Equation~\cite{TaheriEur.Phys.J.D2017}, where the reference is phase-modulated:

\begin{widetext}
\begin{align}
    \frac{\partial a(\theta, t)}{\partial t} &= (-\frac{\kappa}{2} + i \delta\omega_\mathrm{pmp})a %
    + i \sum_\mu A(\mu, t) D_\mathrm{int} \mathrm{e}^{i\mu \theta} %
    + i\gamma L |a|^2 a %
    + i\sqrt{\kappa_\mathrm{ext} P_\mathrm{pmp}} \nonumber\\
    &+  i\sqrt{\kappa_\mathrm{ext}P_\mathrm{ref}} \mathrm{exp} \left[i\mu_s \theta  + i \left(\delta_\mathrm{ref} - \delta_\mathrm{pmp} \right)t + i D_\mathrm{int}(\mu_s)t +  iA\sin\left(\widetilde{\Omega}_\mathrm{RF} t \right) \right]
    \label{eq:LLE}
\end{align}
\end{widetext}
with $A(\mu, t) = \int_{-\pi}^{\pi} a(\theta, t) \mathrm{e}^{-i\mu \theta} \mathrm{d}\theta$ the Fourier transform of the intracavity field, $\mu$ the mode number relative to the main pump, $\delta\omega_\mathrm{pmp}$ and $\delta\omega_\mathrm{ref}$ the detuning of the main and reference pump relative to their resonance frequencies, $\gamma = 2.5$~W\textsuperscript{-1}$\cdot$m\textsuperscript{-1} the effective nonlinear coefficient of the microring resonator under study that will be used in the experimental demonstration later, and $L=2\pi \times 23$~{\textmu}m the circumference of the cavity. We choose the same integrated disperison $D_\mathrm{int}$ as in ref.~\cite{MoilleNature2023}, where higher-order dispersion enables zero-crossings for which phase matching with the DKS occurs. This helps create so-called dispersive waves (DWs), where the comb tooth is on resonance with the cavity mode, increasing both the comb tooth power and the intracavity field of the reference in the KIS regime. Although this DW is not indispensable for KIS to be triggered~\cite{WildiAPLPhotonics2023}, it increases by orders of magnitude the KIS efficiency, leading to larger synchronization bandwidth, hence lower $\beta$, while operating at low reference pump power, and higher $\Omega_\mathrm{RF}$ operation for our study. For this theoretical study,  we chose $\frac{1}{2}\kappa=\kappa_\mathrm{ext} = 200$~MHz,  $\mu_s = -90$ where the DW occurs, $P_\mathrm{main} = 200$~mW, and $P_\mathrm{ref} = 1$~mW. This results in $\Omega_0 =209.5$~MHz, which from the Adler simulation lets us chose $\Omega_\mathrm{RF} = 372$~MHz.

We are able to reproduce a similar phase space in $\{\alpha, A\}$ using the effective resistivity $\nicefrac{\partial \Omega}{\partial \alpha}$ by extracting the beat note between the sideband, carrier and comb tooth (see Methods), which confirms the Adler equation model accurately describe the KIS dynamics, as both ISSs and FSSs are evident~[\cref{fig:3}a]. We note an asymmetry in the mapping results compared to the LLE because of the comb DW not being exactly on resonance. Hence, a slight offset between the comb tooth and the center of the resonance frequency exists, where the maximum of the intracavity reference energy happens at $\alpha\neq0$. The main aspects of the LLE in comparison to the Adler equation -- at the expense of a much more computationally intense calculation -- is the ability to capture significant more physical insight about the dynamics of the system. In particular, one can create a spectrogram out of the interference pattern at $\mu_s$ between the comb tooth and the different waves at a different phase [\cref{fig:3}b, see Methods for details]. The carrier and sideband detunings are clearly visible along with their synchronization windows where their corresponding beat notes vanish. The FSSs are also evident when the beat with the carrier/sideband and comb tooth locks to a rational fraction. Interestingly, other beats with integer multiple slopes and different sign are also visible, corresponding to the new wave at a different frequency from the four-wave mixing (FWM) between the carrier, sidebands, and comb tooth in the system. 

\begin{figure*}
    \begin{center}
        \includegraphics{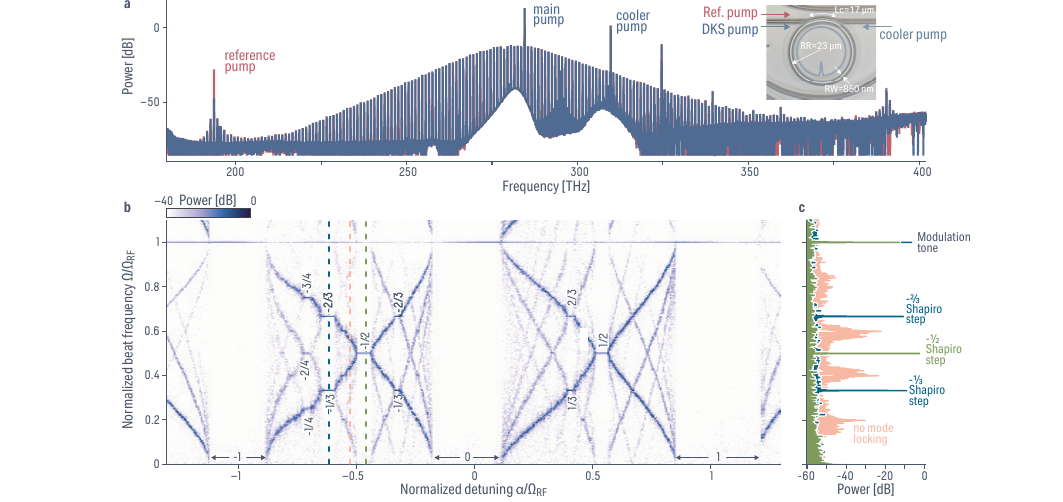}
        \caption{\label{fig:4}%
        \textbf{Experimental demonstration of the FSSs in KIS -- } %
        \textbf{a} Experimental DKS frequency comb obtained from the $R=23$~{\textmu}m \ce{Si3N4} microring resonator presented in the inset, without (blue) and with the reference pump (red). The different laser injection directions are depicted by the arrows in the inset. %
        \textbf{b} Spectrogram obtained from the beat note at the reference mode $\mu_s$, obtained in the same fashion as the LLE results presented in~\cref{fig:3}b with a modulation strength A = 0.75. The ISSs and FSSs are highlighted by their respective synchronization order. Similar to \cref{fig:3}, the offset between the comb tooth and center of the resonance frequency explains the asymmetry in the beat-map, which in this case is of opposite sign from the simulation because of a slight variation in the repetition rate (less than one percent) between simulation and experiment. %
        \textbf{c} Spectra extracted from the spectrogram at their corresponding dashed line colors in \textbf{b}, highlighting the narrow beat note from the mode-locking in the synchronization regime compared to out-of-sync, where the beat notes are much broader. Throughout the experiment, all the lasers are free-running. 
        }
    \end{center}
\end{figure*}
\vspace{1em}


To understand the FSSs in the KIS context it is important to recall the intrinsic $\chi^{(3)}$ nonlinearity on which KIS relies (more precisely, cross-phase modulation) and that FWM can arise out of synchronization. From the second order derivative nature of the Adler equation describing our system, we know that that higher harmonics of the reference detuning $\alpha$ play an important role~\cite{ValizadehJNMP2008}, here created through the FWM interaction. We consider the $\nicefrac{1}{2}$ FSS synchronization window, where the reference carrier and sideband are close to equally surrounding the microcomb tooth at a detuning of $\pm \frac{1}{2}\Omega_\mathrm{RF}\pm \epsilon$, with $\epsilon$ being a small detuning within the FSS synchronization bandwidth [\cref{fig:3}d.i]. Optical parametric amplification (OPA) via stimulated four-wave mixing  between the carrier (pump) and comb tooth (signal) creates a new idler~\cite{MoillearXiv2023} at a frequency $\Omega_\mathrm{RF} + 2\epsilon$ detuned from the comb tooth [\cref{fig:3}d.ii], which is observed in the LLE beat-map results in~\cref{fig:3}b, since its detuning is always twice that of the reference carrier pump. Other FWM effects can also occur, in particular Bragg Scattering (FWM-BS) where the carrier and sideband reference act as pumps that enable noiseless frequency translation at their frequency separation, in this case always defined by $\pm\Omega_\mathrm{RF}$. It allows frequency translation of the OPA idler by $-\Omega_\mathrm{RF}$, which in this case can lie in the KIS bandwidth, capturing the comb tooth [\cref{fig:3}d.iii] even if power is relatively low, since DC-KIS has been demonstrated at below 100~{\textmu}W of in-waveguide power to trigger synchronization. In addition, the OPA idler will be more powerful the closer to the resonance it is created, hence this effect is effective for small RF modulation frequency, consistent with a cavity linewidth of about 400~MHz at the synchronization mode. This translated tone is also visible in~\cref{fig:3}(b) and its beat note vanishes when entering the FSS KIS. Cascaded OPA idlers are created by cascaded FWM, and are also visible in the LLE beat-map. If such an idler is detuned from the comb tooth by close to $\Omega_\mathrm{RF}$, it can be FMW-BS translated close enough to the comb tooth to trigger KIS at another Farey fraction of the carrier detuning. As the cascaded idlers become increasingly less powerful, the KIS energy for synchronization is progressively reduced the higher the order of cascading, reducing the KIS bandwidth of the FSS accordingly with the Farey sequence.  For a more detailed highlight of each nonlinear process and its correspondence on the beat-map, we refer to~\cref{sec:sup_nlodetail}. 
The repetition rate entrainment slope for all of the ISSs or FSSs follows the same optical frequency division factor against the reference detuning $\alpha$, defined by the main pump and the mode at which KIS occurs, in our simulation $\mu_s= -90$ [\cref{fig:3}c]. Despite the OPA idler detuning following a multiple of the reference detuning [\cref{fig:3}d.ii], once in the FWM-BS mediated KIS regime corresponding to the FSSs, the comb tooth being captured also experiences a detuning. This new comb within the $\alpha$-space dimension (in opposition to the azimuthal mode $\mu$-space since all these synchronization phenomena occur at the same mode $\mu_s$) ends up being mode-locked and hence its free spectral range (FSR) must be constant despite any detuning of the reference being entirely absorbed in the $\mu$-space repetition rate, similar to DC-KIS.

\begin{figure}[!t]
    \begin{center}
        \includegraphics{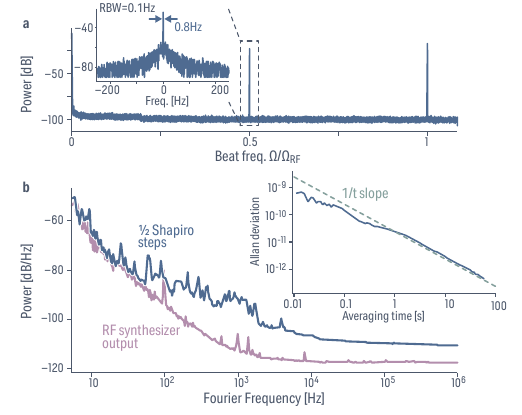}
        \caption{\label{fig:5} 
        \textbf{Experimental demonstration of mode-locking in the AC-KIS regime -- } %
        \textbf{a} Electrical power spectrum of the beat note in the 1/2 FSS regime, with the $x$-axis normalized by $\Omega_\mathrm{RF} = 387.333$~MHz. The inset is a zoom-in on the carrier/comb tooth beat note at $\nicefrac{\Omega_\mathrm{RF}}{2}$, with a resolution bandwidth of 0.1~Hz, highlighting the linewidth of about 0.8~Hz, while every laser in the experiment is free running.  %
        \textbf{b} Phase noise of the $\nicefrac{1}{2}$ Shapiro step mode-locked beat note between the carrier and reference (blue) with all lasers free-running. For reference, the RF synthesizer noise, which is expected to set the noise floor, is plotted (purple). Additional noise comes from unmitigated impedance mismatch in the setup. The power is referenced to that of the carrier, namely, dBc/Hz. Allan deviation (inset) of the same beat note using the RF synthesizer reference as a clock, highlighting the synchronization of the mode-locking with the phase modulation signal as confirmed by its $1/t$ trend (dashed line). %
        }
    \end{center}
\end{figure}
\vspace{1em}
\textbf{Experimental KIS fractional Shapiro steps and their mode-locking -- }
Finally, we aim to experimentally demonstrate the predictions of the previous sections, including the appearance of FSSs (Adler and LLE models) and mode-locking in the $\alpha$-space (LLE model). Our experimental system is a ring width $RW=870$~nm microring resonator made of $H=670$~nm thick \ce{Si3N4} embedded in \ce{SiO2}, similar to the one in ref~\cite{MoilleNature2023}. We create a fundamental transverse electric mode DKS~[\cref{fig:4}(a)] by pumping it at 283~THz with about 150~mW of main pump power on-chip, while thermally stabilizing the resonator with a cross-polarized and counterpropagating high-power cooler pump for adiabatic access to the DKS~\cite{MoilleNat.Commun.2021a,ZhangOptica2019a, ZhouLightSciAppl2019}. We use a reference pump with about 600~{\textmu}W on-chip power aligned with the dispersive wave comb tooth of the DKS around 193~THz. We modulate the reference using a commercially available electro-optic phase modulator at 387.333~MHz with 158~mW power, corresponding to a modulation strength of $A=0.75$, and sweep the reference over about 1~GHz at a 70~Hz rate using a ramp signal controlling the piezo-electric element of the continuously tunable reference pump laser. We record the beat note between the reference and the nearest comb tooth using a 6~GHz fast photodiode, which we process using a fast oscilloscope. Similar to the processing of the LLE simulations, such temporal beat note traces allow us to reconstruct a spectrogram as shown in~\cref{fig:4}(b), whose behavior is in good overall agreement with the LLE result~[\cref{fig:3}(b)], with the emergence of both ISSs and FSSs clearly visible. Processing the spectrograph in the FSSs regime~[\cref{fig:4}c] highlights the mode-locking phenomenon. The FSS beat notes that appear in KIS have narrow linewidth, despite all of the lasers being free-running (each with a linewidth of $\approx 500$~kHz). In contrast, out-of-KIS, beat notes are broad, due to the optical linewidth of the relevant comb tooth being at least $\mu_s$ times larger (\textit{i.e.,} about $\Omega_\mathrm{RF}/10$). 

We further study the $\nicefrac{1}{2}$ Shapiro step mode-locking regime. Measuring a similar spectrum~[\cref{fig:5}(a)] as in~\cref{fig:4}c with a real-time spectrum analyzer highlights the narrow linewidth of the carrier/comb tooth beat note of about 0.8~Hz, which is many orders-of-magnitude lower than either the comb tooth or the reference pump linewidth. Phase-noise analysis of this beat note produces a result similar to the noise of the RF synthesizer that is driving the phase modulator~[\cref{fig:5}(b)], as expected from the mode-locking physics described previously (and with more details in~\cref{sec:sup_modlock}), for which only FWM-BS enables noiseless frequency conversion~\cite{LiNaturePhoton2016, SinghOptica2019} from the reference onto the captured comb tooth. This holds true within the AC-KIS mode-locking since the carrier and sideband mediated by FWM-BS share common noise and as a result the beat note should only carry the RF synthesizer noise. In addition, the Allan deviation of the beat note clocked by the RF synthesizer reference~[\cref{fig:5}(b), inset] shows a clear $1/t$ trend, demonstrating synchronization of the sub-comb with the phase modulation RF signal, and highlighting further the FWM-BS induced mode-locking effect.

\vspace{1em}
\textbf{Discussion -- } In conclusion, we have harnessed the similarities between different physical systems that follow the same second-order Adler equation as Kerr-induced synchronization in a microring resonator to predict the existence of soliton frequency comb analogues to effects that occur in Josephson junctions, such as integer and fractional Shapiro steps. We show that the predictions of the Adler equation, which is a general model for synchronization, are reproduced by the LLE commonly used to model soliton microcombs, thereby validating use of the Adler equation in the context of KIS. Although the Adler model correctly predicts effects that we reproduce accurately experimentally using an octave-spanning integrated microcomb, the LLE provides deeper physical insight into the observed effects, particularly the optical fractional Shapiro steps. We find that these fractional steps are explained through a combination of different four-wave mixing processes, which further explain the mode-locking of the sub-comb that is formed by the frequency mixing of the comb tooth and reference pump. Through noise and stability measurements, we experimentally demonstrate such mode-locking. Beyond introducing a new method to synchronize and externally control a microcomb, which can find direct application in metrology and microwave generation, it is important to note that the AC-KIS effect is created from a phase modulation that can either be on the reference pump laser or on the DKS. Here, we have chosen the former for its convenience as an external tunable parameter. Going forward, synchronization to other Kerr microcomb states, such as breather solitons~\cite{puzyrev_frequency_2022}, can be expected to exhibit similar behavior, while also potentially displaying important new features such as a coherent link between the soliton repetition rate and breather frequency. Finally, the demonstration of analogous behavior to superconducting circuits within a room-temperature, commercially-fabricated photonic platform points to the possibility of using photonics to understand and engineer synchronization phenomena in condensed matter physics.

                          
\bibliographystyle{naturemag}

\clearpage

\noindent \textbf{\Large Methods}

\vspace{1em}
\noindent \textbf{LLE repetition rate extraction}\\
The LLE presented in~\cref{eq:LLE} is in the azimuthal coordinate of the resonator ($\theta$) moving at at a fixed speed $D_1$ such that $\theta = \tilde{\theta} - D_1t$, with $\tilde\theta$ the resonator angle in the frame of the laboratory. $D_1$ is chosen to be as close as possible to the repetition rate of the DKS $\omega_\mathrm{rep}$ to remove any fast oscillation in the mean field approximation. By sampling the LLE at every time step $2\pi/D_1$, the angular position of the DKS can be retrieved. If the position is fixed, $\omega_\mathrm{rep} = D_1$, otherwise, the repetition rate can be easily retrieved through $\omega_\mathrm{rep} = D_1  - \frac{\partial \theta_0(t)}{\partial t}$, where $\theta_0(t)$ is the angular position of the soliton peak with time.

\vspace{1em}
\noindent \textbf{Phase space reconstruction}\\
For both the LLE simulation and experiment, the same metrics are recorded. In the experiment, the beat note between the reference and the DKS is measured with a fast photodiode for which we record the temporal trace against the detuning of the reference. For the simulation, we solve~\cref{eq:LLE} with an adiabatic linear detuning ramp of the reference $\delta \omega_\mathrm{ref}$ over 100 million round trips. We can extract the field $A(\mu, t) = \int_{-\pi}^{\pi} a(\theta, t)\mathrm{exp}\left( -i\mu\theta \right) \mathrm{d}\theta$ and only focus on $A\left( \mu_\mathrm{s},t \right)  = \sum_k |A_k (\mu_s)|\mathrm{e}^{i\delta \omega_k}$, which carries all the amplitude and phase of the waves at the same azimuthal mode and with different phase velocity, here described by the $k$ subscript. Hence, similar to an interferometer, extracting $\left|A(\mu_s)\right|^2$ results in a modulated signal, identical to the experiment conducted, from which we could process the spectrogram. Using the \texttt{scipy.signal} package with a Hanning window function, we retrieve the two-dimensional mapping in time and offset frequency from the comb tooth of the different colors in the system, resulting in the plot and interpretation presented in~\cref{fig:3,fig:4}. For the effective resistivity map in~\cref{fig:3}a, only the beat between the comb tooth and reference carrier signal is extracted, and the slope for each modulation strength $A$ displayed. 

\vspace{2em}
\noindent \textbf{\Large Acknowledgments} \\
The photonic chips were fabricated in the same fashion as the ones presented in~\cite{MoilleNature2023}, based on a thick Si$_3$N$_4$ process in a commercial foundry~\cite{rahim_open-access_2019}.

\noindent The Scientific color map batlow~\cite{Crameri2023} is used in this study to prevent visual distortion of the data and exclusion of readers with colour-vision deficiencies~\cite{CrameriNatCommun2020}. 

\noindent We acknowledge partial funding support from the Space Vehicles Directorate of the Air Force Research Laboratory, the Atomic–Photonic Integration programme of the Defense Advanced Research Projects Agency, and the NIST-on-a-chip program of the National Institute of Standards and Technology. P.S and C.M. acknowledges support from the Air Force Office of Scientific Research (Grant No. FA9550-20-1-0357) and the National Science Foundation (Grant No. ECCS-18-07272). We thank Sashank Sridhar, Jordan Stone, and Sean Krzyzewski for insightful feedback. Certain commercial products or names are identified to foster understanding. Such identification does not constitute recommendation or endorsement by the National Institute of Standards and Technology, nor is it intended to imply that the products or names identified are necessarily the best available for the purpose. 

\vspace{1em}
\noindent \textbf{\Large Author Contributions} \\
G.M. and K.S. led the project. G.M. designed the resonators, developed the Adler model and performed the measurements and simulations. P.S. and C.M. helped with the theoretical and numerical results. U.A.J. and M.C. helped with the experimental setup. G.M. and K.S. wrote the manuscript, with input from all authors. All the authors contributed and discussed the content of this manuscript.

\vspace{1em}
\noindent \textbf{\Large Competing Interests} \\
G.M., C.M. and K.S have submitted a provisional patent application based on aspects of the work presented in this paper.

\vspace{1em}
\noindent \textbf{\Large Data availability} \\
The data that supports the plots within this paper and other findings of this study are available from the corresponding authors upon request.

\vspace{1em}
\noindent \textbf{\Large Code availability} \\
The simulation code is available from the authors through the pyLLE package available online~\cite{MoilleJ.RES.NATL.INST.STAN.2019}, with modifications that are availabe upon reasonable request, using the inputs and parameters presented in this work.
\vfill
\clearpage
\beginsupplement

                                  
\section{\label{sec:sup_parameter} Adler equation parameters and relation with other similar systems}

Many systems follow the same Adler equation that has been used in this work. A key aspect of our work is that, since systems which may seem far apart follow the same Adler equation, one could learn from one of these systems and transfer that understanding onto another system. In our case, the AC Josephson junction (JJ) with integer and fractional Shapiro steps can also be observed in Kerr-induced synchronization (KIS). It is therefore interesting to note how each system is normalized to obtain~\cref{eq:DC_Adler,eq:AC_Adler}, here written again for the convenience of the reader as \cref{eq_sup:DC_Adler,eq_sup:AC_Adler}: 

\begin{align}
    \beta \frac{\partial^2 \Phi}{\partial \tau^2} +  \frac{\partial \Phi}{\partial \tau} &= \alpha + \sin(\Phi)
    \label{eq_sup:DC_Adler}\\
    \beta \frac{\partial^2 \Phi}{\partial \tau^2} + \frac{\partial \Phi}{\partial \tau} &= \alpha + \sin(\Phi) + A\sin(\Omega_\mathrm{RF}\tau)
    \label{eq_sup:AC_Adler}
\end{align}

We have summarized these parameters for the damped pendulum under constant torque, the JJ, and the KIS in~\cref{tab:sup_param}. 

\begin{table}[H]
    \captionsetup{width=1\linewidth}
    \caption{\label{tab:sup_param}\textbf{Physical parameters used in the Adler equation --} The parameters are extracted from the LLE and the experiment.}
    \renewcommand{\arraystretch}{3}
    \begin{center}
        \begin{tabularx}{1\linewidth}{c Y Y Y}    
            \toprule
              & Damped pendulum under constant torque~\cite{CoulletAmericanJournalofPhysics2005} & Josephson junction~\cite{ValizadehJNMP2008} & Kerr-induced synchronization~\cite{MoilleNature2023}\\
            \midrule
            $\displaystyle \tau = \Omega_0 t$  %
                & %
                $\displaystyle \frac{mgl}{v}$
                & %
                $\displaystyle \frac{2e}{\hbar}I_\mathrm{c} Rt$
                & $\displaystyle  2\mu_s  D_1 E_\mathrm{kis} t$ \\
            $\displaystyle \Gamma$  %
                & %
                $\displaystyle \frac{mgl}{I}$
                & %
                $\displaystyle \frac{1}{RC}$
                & $\kappa$ \\
            $\displaystyle \beta = \frac{\Omega_0}{\Gamma}$  %
                &  %
                $\displaystyle \frac{I}{v} $
                & %
                $\displaystyle  \frac{2 e}{\hbar}I_cR^2 C$
                & $\displaystyle  \frac{2\mu_s  D_1 E_\mathrm{kis}}{\kappa} $ \\
            $\displaystyle\alpha = \frac{\Omega_\mathrm{ext}}{\Omega_0}$  %
            &  %
            $\displaystyle \frac{v}{mgl}\Omega_m$ 
            &  %
            $\displaystyle \frac{I_\mathrm{ext}}{I_\mathrm{c}}$
            & %
            $\displaystyle \frac{\delta\omega_\mathrm{ref} + D_\mathrm{int}(\mu_s) + \delta\omega_\mathrm{NL}}{2\mu_s  D_1 E_\mathrm{kis}}$\\
            \bottomrule
        \end{tabularx}
    \end{center}
    \renewcommand{\arraystretch}{1}
\end{table}

\noindent In the above table, for the damped driven pendulum $m$ is the pendulum mass, $l$ is its length, $g$ is the gravitational acceleration, $I$ is the moment of inertia, $v$ is the viscous torque, and $\Omega_m$ is the applied rotation speed of the driving motor. For the JJ in the resistively and capacitively shunted junction (RCSJ)  model, $C$ is the capacitance, $R$ is the resistance, $I_\mathrm{c}$ is the critical current, $I_\mathrm{ext}$ is the external current, $\hbar$ is the reduced Planck constant, and $e$ is the electron charge. %
We note that the parameter $\beta$ used here in the JJ under the RCSJ model is known as the Stewart-McCumber parameter~\cite{BlackburnPhysicsReports2016}. For the KIS, $\mu_s$ is the azimuthal mode at which synchronization occurs normalized to the fixed main pump, $\kappa$ is the cavity decay rate with $\kappa = 2\kappa_\mathrm{ext}$ ($\kappa_\mathrm{ext}$ is the waveguide coupling rate), $D_1$ is the free spectral range without synchronization, $E_\mathrm{kis} = \sqrt{\frac{\kappa_\mathrm{ext}}{\kappa^2}P_\mathrm{ref}E_\mathrm{\mu s}}/E_\mathrm{dks}$ is normalized KIS coupling energy with $E_\mathrm{dks}=\int |a|^2 \mathrm{d}\theta/2\pi$, $P_\mathrm{ref}$ is the reference pump power, $\delta\omega_\mathrm{ref}$ is the detuning of the reference pump, $D_\mathrm{int}(\mu_s)$ is the integrated dispersion at the KIS mode, and $\delta\omega_\mathrm{NL} = D_\mathrm{int}(\mu_s) + \mu_sD_1(\mu_0)(1 + 2E_\mathrm{NL} - 2E_{0})$ is the nonlinear detuning experienced by the $\mu_s$ comb tooth, with $E_{0}$ and $E_\mathrm{NL}$ being the cross-phase modulation energy from the whole comb and the main pump, respectively.

Physically, it is interesting to see the resemblance of the normalization and gain some physical insight for each of the systems and analogous behavior between them. $\Omega_0$ is the natural frequency of the oscillator that defines the bandwidth of the synchronization and $\nicefrac{1}{\Gamma}$ is the energy loading time for the system before reaching steady-state. It is particularly apparent in the JJ system, given the $\nicefrac{1}{RC}$ behavior, and in the KIS system is the resonance linewidth at the synchronized mode and hence the rate at which energy can be injected in the cavity. The normalized frequency offset between the leader and follower oscillator $\alpha$ is the ODE linear loss/gain term, compensating for the cosine potential $V=\cos(\Phi)$ for the Adler equation rewritten through a potential term $\partial V/\partial \tau = \sin(\Phi)$~\cite{CoulletAmericanJournalofPhysics2005}.

                                             
\section{\label{sec:sup_modlock} Mode-locking through OPA and FWM-BS}

The mode-locking happening at the fractional Shapiro steps is explained through a two-step process. Here, we take the example of the \nicefrac{1}{2} Shapiro step, but it could be easily extended to the other fractional Shapiro steps. The first step is the creation of an idler from the optical parametric amplification (OPA) of the comb tooth by the reference pump, which also generates an idler. For the sake of simplicity -- since the focus of this section is the phase noise of each component at play -- we dismiss the linear part, as it does not play a role in the wave phase. The idler OPA follows: 
\begin{equation}
    -\frac{i}{\gamma L}\frac{\partial a_\mathrm{idl}}{\partial t} = a_\mathrm{ref}^2 a_\mathrm{dks}^* \\ 
\end{equation}

\noindent where $\gamma$ is the effective Kerr nonlinearity, $L$ is the resonator circumference, and $a_\mathrm{(idl;ref;dks)}$ is the intracavity field amplitude for the idler/reference/DKS, respectively.

The second effect is the spectral translation by $-\Omega_\mathrm{RF}$ of the idler by the carrier and sideband reference pump through four-wave mixing Bragg scattering (FWM-BS). Assuming that this FWM-BS translated idler is capturing the comb tooth $a_\mathrm{dks}$ lets us write in this scenario:

\begin{equation}
    -\frac{i}{\gamma L}\frac{\partial a_\mathrm{dks}}{\partial t} = a_{sb} a_\mathrm{ref}^* a_\mathrm{idl}
\end{equation}

While this two-step effect helps with the physical understanding of the system, in reality they do happen simultaneously once KIS is achieved such that: 

\begin{align}
    \frac{i}{\gamma L}\frac{\partial a_\mathrm{dks}}{\partial t} &= a_\mathrm{ref}\mathrm{e}^{-i\frac{\Omega_\mathrm{RF}}{2}t}\left( a_\mathrm{ref}a_\mathrm{idl}^* \mathrm{e}^{i\frac{\Omega_\mathrm{RF}}{2}t} + a_\mathrm{ref}^* a_\mathrm{idl} \mathrm{e}^{-i\frac{\Omega_\mathrm{RF}}{2}t}  \right)\nonumber\\
    & = \Gamma a_\mathrm{ref}\mathrm{e}^{-i\frac{\Omega_\mathrm{RF}}{2}t}
\end{align}

\noindent with $\Gamma = \mathfrak{Re}\left( a_\mathrm{ref}a_\mathrm{idl}^* \mathrm{e}^{i\Omega_\mathrm{RF}t} \right)$ being a real number and hence does not affect the phase of the comb tooth $a_\mathrm{dks}$. The phase of the comb tooth is thus defined as $\varphi_\mathrm{dks} = \varphi_\mathrm{ref}- \varphi_\mathrm{RF}$ with $\varphi_\mathrm{RF} = \Omega_\mathrm{RF}t$. Accounting for phase/frequency noise in both the reference pump $\delta\varphi_\mathrm{ref}$ and microwave synthesizer $\delta\varphi_\mathrm{RF}$, the comb tooth now carries a noise $\delta\varphi_\mathrm{dks} = \delta\varphi_\mathrm{ref} - \frac{\delta\varphi_\mathrm{RF}}{2}$. Hence, despite the reference and comb tooth being offset from one another, it highlights their mode locking since they have become phase-coherent in the FSS condition.

The experimental beat note does not discriminate between the carrier-comb tooth beat and the sideband-comb tooth beat, and hence the noise is accounted for at least twice and sums up such that $\delta\varphi_\mathrm{beat}\equiv \delta\varphi_\mathrm{RF}$, which is consistent with the results in~\cref{fig:5} that show that the RF synthesizer noise sets the beat note noise floor and long-term stability.

\section{\label{sec:sup_setup} Experimental setup}
\begin{figure}[htbp]
    \centering
    \includegraphics{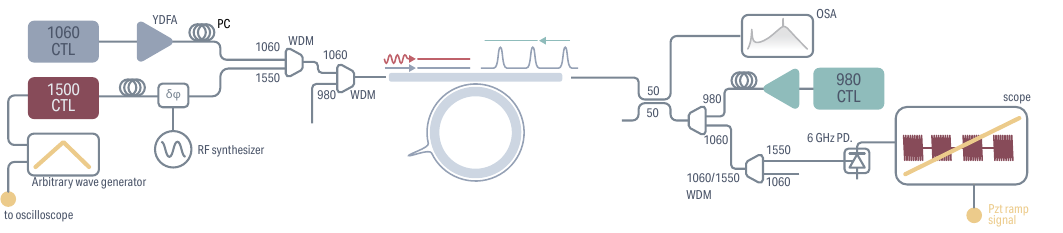}
    \caption{\label{sup_fig:setup} Experimental setup for DKS generation and KIS. CTL: continuously tunable laser; YDFA: ytterbium-doped fiber amplifier; PC: polarization controller; WDM: wavelength demulitplexer; OSA: optical spectrum analyzer; PD: photodiode; PZT: piezo-electric transducer.}
\end{figure}
The experimental setup is depicted in~\ref{sup_fig:setup}. For the sake of simplicity, we have not depicted basic power detection probes and monitors that allow us to verify proper optical injection/collection to the chip nor the probing to verify the DKS state. We pump the fundamental transverse electric mode of a $23$~{\textmu}m \ce{Si3N4} microring resonator using a 1060~nm continuously tunable laser (CTL) with a pump frequency of about 283~THz and with about 150~mW on-chip power, obtained thanks to a ytterbium-doped fiber amplifier (YDFA). We use a counterpropagating and cross-polarized cooler pump from a 980~CTL at about 307~THz to thermally stabilize the cavity, which enables us to reach the DKS state with a simple adiabatic frequency tuning of the main pump. We phase modulate the 1550~nm reference CTL   using an electro-optics modulator driven by a microwave synthesizer. The piezo element of the 1550 CTL is controlled by a ramp signal from an arbitrary waveform generator that enables the frequency tuning of the reference. We combine the reference and the main pump using wavelength demultiplexers (WDMs).

                                                
\section{\label{sec:sup_nlodetail} Detailed beat-map to explain the four-wave mixing Bragg-scattering mediated Kerr-induced synchronization}

\begin{SCfigure}
    \includegraphics[width = 0.75\textwidth]{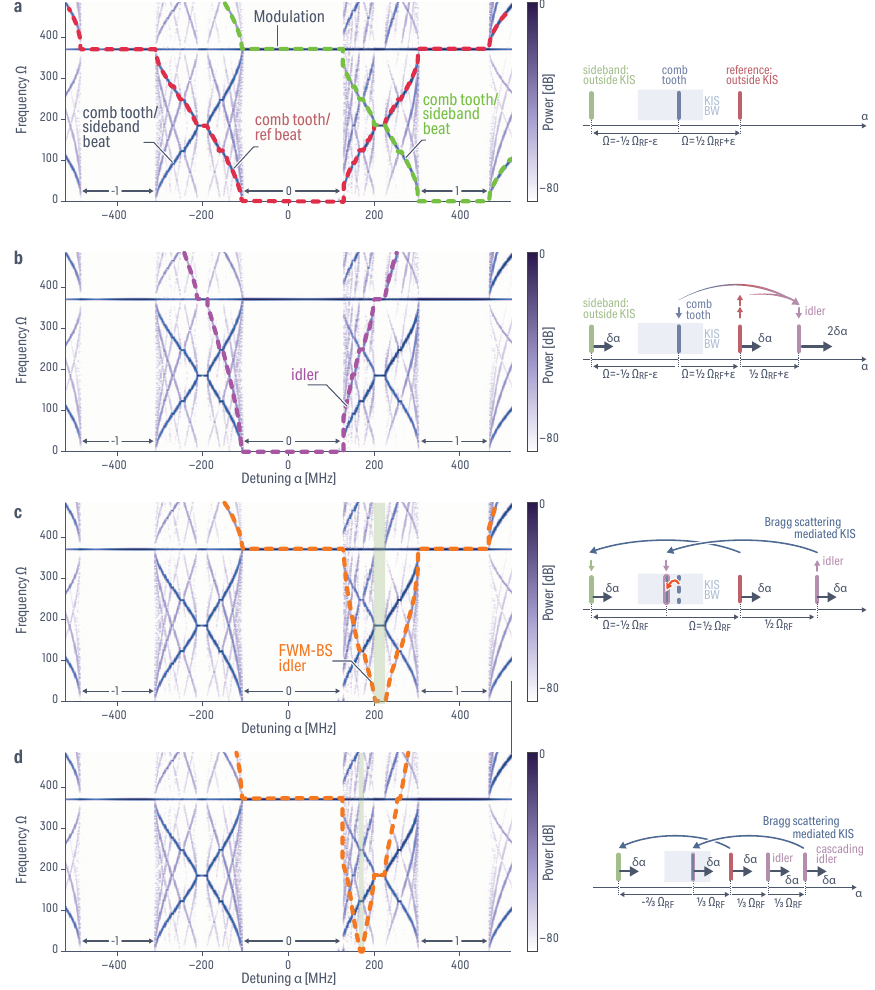}
    \caption{\label{sup_fig:modeLockingExplain}
    \textbf{Detailed explanation of the different beat notes and their association with their respective colors -- } It is important to note that only the absolute value of the beat can be recorded in experiment. 
    \textbf{a} The reference carrier is detuned at a rate $\delta\alpha$ (red) and its sideband follows the same detuning rate shifted by $\Omega_\mathrm{RF}$. Each of them independently creates a KIS window highlighted by the absence of a beat other than the modulation one. 
    \textbf{b} The nonlinear mixing between the comb tooth and the reference carrier enables an optical parametric amplifier (OPA) idler to be created, which per energy conservation will be detuned at twice the rate of the carrier reference ($2\delta\alpha$), as highlighted in purple. 
    \textbf{c} The carrier and sideband reference enables the creation of a nonlinear grating, which through four-wave mixing Bragg scattering, can frequency translate the idler by $\Omega_\mathrm{RF}$ (orange). Once this FWM-BS idler comes close enough to the comb tooth at $\Omega=0$, it can trigger the Kerr-induced synchronization, as highlighted in the green shaded area. This corresponds to the half-fractional Shapiro steps where a constant beat between the reference and the comb tooth is observed at $\Omega_\mathrm{RF}/2$. 
    \textbf{d} The same process can be repeated for a cascaded OPA idler, explaining the other fractional Shapiro steps.
    }
\end{SCfigure}

To simplify the comprehension of~\cref{fig:3}b and the physical process involved in the four-wave mixing Bragg-scattering (FWM-BS) mediated Kerr-induced synchronization at the core of the fractional Shapiro steps (FSSs), we break down the beat-map to highlight the contribution of each color and their significance [\ref{sup_fig:modeLockingExplain}].

\end{document}